\documentclass[twocolumn,aps,prl,superscriptaddress,nofootinbib,10pt]{revtex4-2}
\usepackage{amssymb}
\usepackage{amsmath,bm}
\usepackage{graphicx}
\usepackage[normalem]{ulem}
\usepackage{multirow}
\usepackage[colorlinks,linkcolor=blue,urlcolor=blue,citecolor=blue]{hyperref}
\usepackage{lipsum}
\usepackage[usenames, dvipsnames]{xcolor}
\usepackage{tensor}
\usepackage{isotope}
\usepackage{amsmath}
\usepackage{color}
\usepackage{bm}
\allowdisplaybreaks[4]
 
\renewcommand{\sout}{\bgroup \color{red} \ULdepth=-.5ex \ULset}

\usepackage{tikz,xcolor,hyperref}
\definecolor{lime}{HTML}{A6CE39}
\DeclareRobustCommand{\orcidicon}{
	\begin{tikzpicture}
	\draw[lime, fill=lime] (0,0) 
	circle [radius=0.16] 
	node[white] {{\fontfamily{qag}\selectfont \tiny ID}};
	\draw[white, fill=white] (-0.0625,0.095) 
	circle [radius=0.007];
	\end{tikzpicture}
	\hspace{-2mm}
}

\foreach \x in {A, ..., Z}{%
	\expandafter\xdef\csname orcid\x\endcsname{\noexpand\href{https://orcid.org/\csname orcidauthor\x\endcsname}{\noexpand\orcidicon}}
}

\begin{document}

\title{Alpha clustering in warm and dense nuclear matter from heavy-ion collisions}

\author{Rui Wang\orcidA{}}
\thanks{Present address: INFN, Laboratori Nazionali del Sud, I-$95123$ Catania, Italy}
\email{rui.wang@lns.infn.it}
\affiliation{Key Laboratory of Nuclear Physics and Ion-beam Application~(MOE), and Institute of Modern Physics, Fudan University, Shanghai $200433$, China}
\affiliation{Shanghai Institute of Applied Physics, Chinese Academy of Sciences, Shanghai $201800$, China}
\affiliation{Istituto Nazionale di Fisica Nucleare (INFN), Sezione di Catania, I-$95123$ Catania, Italy}

\author{Zhen Zhang\orcidC{}}
\email{zhangzh275@mail.sysu.edu.cn}
\affiliation{Sino-French Institute of Nuclear Engineering and Technology, Sun Yat-Sen University, Zhuhai 519082, China}

\author{Yu-Gang Ma\orcidB{}}
\email{mayugang@fudan.edu.cn}
\affiliation{Key Laboratory of Nuclear Physics and Ion-beam Application~(MOE), and Institute of Modern Physics, Fudan University, Shanghai $200433$, China}
\affiliation{Shanghai Research Center for Theoretical Nuclear Physics, NSFC and Fudan University, Shanghai $200438$, China}

\author{Lie-Wen Chen\orcidD{}}
\email{lwchen@sjtu.edu.cn}
\affiliation{State Key Laboratory of Dark Matter Physics, Key Laboratory for Particle Astrophysics and Cosmology (MOE), and Shanghai Key Laboratory for Particle Physics and Cosmology, School of Physics and Astronomy, Shanghai Jiao Tong University, Shanghai 200240, China}

\author{Che Ming Ko\orcidE{}}
\email{ko@comp.tamu.edu}
\affiliation{Cyclotron Institute and Department of Physics and Astronomy, Texas A\&M University, College Station, Texas 77843, USA}

\author{Kai-Jia Sun\orcidF{}}
\email{kjsun@fudan.edu.cn}
\affiliation{Key Laboratory of Nuclear Physics and Ion-beam Application~(MOE), and Institute of Modern Physics, Fudan University, Shanghai $200433$, China}
\affiliation{Shanghai Research Center for Theoretical Nuclear Physics, NSFC and Fudan University, Shanghai $200438$, China}

\date{\today}
\begin{abstract}
Although light nuclear clusters are known to affect the properties of warm and dilute nuclear matter, their role in warm and dense nuclear matter remains unclear due to the lack of experimental evidence for their modifications by the Mott effect in such an environment. 
To address this issue, we resort to intermediate-energy heavy-ion collisions, where light clusters are mainly produced in the transiently formed warm and dense matter.
A kinetic approach, which includes dynamically the formation and dissociation of light clusters, is employed to deduce the strength of the Mott effects and the $\alpha$-particle fraction in warm and dense nuclear matter from the light-nuclei yields measured by the FOPI Collaboration in central Au$+$Au collisions at energies of $0.25A$ to $0.6A~\rm GeV$.
We find an unexpectedly abundant $\alpha$ clustering in this environment, which will have profound implications for modeling the nuclear equation of state and describing supernovae and neutron star mergers.
\end{abstract}

\pacs{12.38.Mh, 5.75.Ld, 25.75.-q, 24.10.Lx}
\maketitle

\emph{Introduction.}{\bf ---}
It is generally thought that dense nuclear matter can be regarded almost as a uniform liquid composed solely of nucleons.
Such an assumption is based on the belief that the Mott effect~\cite{MotRMP40}, which originates from the Pauli blocking of nucleons in the phase space, weakens the attractive correlations among nucleons that lead to bound-state formation~\cite{RopNPA379,RopNPA399,SchAP202}.
These in-medium effects on light clusters are of crucial importance in determining the composition and thermodynamic properties of nuclear
matter~\cite{TypPRC81,RopPRC92,GulPRC92,ZZWPRC95}. Although the influence of the Mott effect generally decreases with increasing temperature, as shown in studies based on the solution of in-medium Schrodinger equations~\cite{SchAP202,RopPRC79}, experimental indications are still lacking on the extent to which light-cluster formation and their abundances are modified in warm and dense nuclear matter.

Empirical analyses based on heavy-ion collisions around the Fermi energy~\cite{BouPRC97}, where light clusters freeze out at rather low densities, have already allowed the determination of in-medium properties of light clusters in warm and dilute nuclear matter~\cite{HagPRL108,QLPRL108,PaiPRL125,CusPRL134}.
For clusters in warm and dense nuclear matter where their properties remain unexplored, one needs to resort to heavy-ion collisions beyond the Fermi energy and up to the intermediate energy in the GeV regime, where the produced nuclear matter can reach (locally) a temperature as high as a few tens of MeV and a density of $2$-$3$ times of the nuclear saturation density ($\rho_0$ $\approx$ $0.16~\rm fm^{-3}$)~\cite{BerPR160,LBANPA708,DXGPRC94}. Light clusters produced in these collisions may therefore freeze out at much higher densities than those in collisions around the Fermi energy. Since light nuclei are abundantly produced in these collisions~\cite{ReiNPA848,OnoPPNP105,BouSmtr13}, their expected suppression in warm and dense nuclear matter may not be as large as usually expected. A quantitative study of this reduced in-medium suppression of light clusters, particularly the possible large $\alpha$-particle abundance, will have great implications in understanding the equation of state of warm and dense nuclear matter~\cite{OerRMP89,Sum2020}, and the physics of supernovae~\cite{PonsApJ513,OerRMP89,Sum2020} and binary neutron-star mergers~\cite{GW170817}, where the nuclear matter can reach similar temperatures and densities~\cite{Sum2020,MosPRD107}. 

For intermediate-energy heavy-ion collisions, where only local---rather than global~\cite{ReiPRL92,AndEPJA30}---equilibrium may be reached, a proper description of light-nuclei production requires dynamical approaches that include light clusters up to at least the $\alpha$ particle and the Mott effect, which is only recently realized in Ref.~\cite{WRPRC108}. 
Here, we employ such an approach to determine the strength of the Mott effect on  deuteron ($d$), triton ($t$), helium-3 ($h$), and the $\alpha$ particle from their measured yields in central Au$+$Au collisions at $0.25A$ to $0.6A~\rm GeV$ by the FOPI Collaboration~\cite{ReiNPA848}. This information then allows for an estimation of light-cluster fractions in warm and dense nuclear matter at temperatures and densities where light clusters undergo chemical freeze-out in these collisions.

\emph{Kinetic approach and Mott effect on light clusters.}{\bf ---}
In the kinetic approach employed in the present study, light clusters, as for nucleons, pions and $\Delta$-resonances, are included dynamically in the kinetic equations that are derived from the real-time Green's function formalism~\cite{DanNPA533,Ram2007}. These kinetic equations, which govern the time evolution of their Wigner functions or phase-space occupations $f_i(\vec{r},\vec{p},t)$, are given as follows,
\begin{equation}\label{E:KE}
 (\partial_t +{\boldsymbol\nabla}_p\epsilon_i\cdot{\boldsymbol\nabla}_r -{\boldsymbol\nabla}_r\epsilon_i\cdot{\boldsymbol\nabla}_p)f_i = I_i^{\rm coll}[f_n,f_p,f_d,\cdots],
\end{equation}
with $i$ denoting the proton ($p$), neutron ($n$), $d$, $t$, $h$ and $\alpha$, as well as the different charged states of pion ($\pi$) and $\Delta$-resonance. In the above equation, $\epsilon_i[f_n,f_p,\cdots]$ represents the single-particle energy, which can be derived from an energy-density functional. 
In this study, it is obtained using an extended Skyrme interaction~\cite{CarPRC78,WRPRC98,WSPPRC109,WSPPRC111}, which is capable of reproducing key astrophysical observables such as the mass-radius relation and the tidal deformability of neutron stars. The collision integral $I_i^{\rm coll}$ consists of a gain term ($<$) and a loss term ($>$),
\begin{equation}\label{E:Ic}
I_i^{\rm coll} = K_i^{<}[f_n,f_p,\cdots](1\pm f_i) - K_i^{>}[f_n,f_p,\cdots]f_i,
\end{equation}
where the plus and minus signs are for bosons and fermions, respectively. Both terms contain contributions from various scattering channels, such as $nnnpp\leftrightarrow n\alpha$, which can be obtained from diagrammatic expansions of many-body Green's functions~\cite{DanNPA533}. 
We note that such a production mechanism of light clusters has been shown to dominate the production of deuteron in nuclear collisions at the LHC energies~\cite{ALIaXv2504}.
The above kinetic equations can be solved using the test-particle method~\cite{WonPRC25}, with the drift terms on the left-hand side treated by the lattice Hamiltonian method~\cite{LenPRC39,WRPRC99}, and the collision integral on the right-hand side treated in a stochastic way~\cite{DanNPA533,WRPLB807,WRFiP8}.  It has been shown that, with the inclusion of the Mott effect, such an approach is able to describe measured light-nuclei yields in central Au+Au collisions at energies of $0.25A$ to $1.0A~\rm GeV$~\cite{WRPRC108}.

The inclusion of in-medium effects on light clusters is essential for a proper dynamical description of their production in heavy-ion collisions.
From the in-medium Schr\"odinger equation, the binding energy of a light cluster $E_{\rm B}(\mathbf{P})$ of center-of-mass momentum $\mathbf{P}$ could vanish, and its constituent nucleons become unbound if the phase-space occupation of nucleons in the nuclear medium around the light cluster is sufficiently large~\cite{RopNPA379,RopNPA399}. This is mainly due to the Pauli blocking of the constituent nucleons inside the light cluster by the surrounding nuclear medium, which weakens their attractive correlations.
A phase-space excluded-volume approach, which captures this key feature of the in-medium Schr\"odinger equation while requires significantly less computational effort, has been introduced in Ref.~\cite{DanNPA533} to solve numerically the kinetic equations. Specifically, the formation of a light cluster of species $\nu$ with mass number $A$ and momentum $\mathbf{P}$ in the nuclear medium is allowed only if the average phase-space occupation of both neutrons and protons around the light cluster is less than a cutoff value $F_{A}^{\rm cut}$, i.e.,
\begin{equation}
\langle f_\tau \rangle_\nu(\mathbf{P}) \equiv \int f_\tau^{\rm tot}(\mathbf{p})|\tilde{\phi}_{\nu,\mathbf{P}}({\bf p})|^2\frac{{\rm d}{\bf p}}{(2\pi\hbar)^3} \leqslant F^{\rm cut}_A.
\label{E:fcut}
\end{equation}
Here, $\tau$ $=$ $n$ or $p$, and $|\tilde{\phi}_{\nu,\mathbf{P}}({\bf p})|^2$ denotes the normalized one-body probability distribution of the nucleons inside the light cluster. In Eq.~(\ref{E:fcut}), the total nucleon occupation $f_\tau^{\rm tot}$ contains contributions from both unbound nucleons and those bound in light clusters, as generalized in Ref.~\cite{WRaXv2506} with respect to the original phase-space excluded-volume approach~\cite{DanNPA533,KuhPRC63,WRPRC108}, where only the former is considered. The parameters $F_A^{\rm cut}$ can be regarded as surrogates of the strength of the Mott effect, with smaller values of $F_A^{\rm cut}$ corresponding to a stronger Mott effect.

For describing heavy-ion collisions, the above criterion for light-cluster formation is incorporated into the kinetic approach through the collision integral in Eq.~(\ref{E:KE}), i.e., the formation of light clusters from many-body scatterings is allowed only if Eq.~(\ref{E:fcut}) is satisfied.
There, the $f_\tau^{\rm tot}$ is obtained from the phase-space occupations $f_i$ in Eq.~(\ref{E:KE}) for both unbound nucleons and light clusters in each spatial lattice and at each time step.

For nuclear matter, Eq.~(\ref{E:fcut})
can be used to define the Mott momentum $P^{\rm Mott}_\nu$, above which a light cluster of species $\nu$ can exist, since
$\langle f_\tau \rangle_\nu(\mathbf{P})$ in the nuclear matter decreases with increasing $|\mathbf{P}|$.
Light clusters thus have the following distributions in nuclear matter,
\begin{equation}
 f_\nu^{\rm eq}(\mathbf{P}) = \frac{H(|\mathbf{P}|-P_\nu^{\rm Mott})}{{\rm exp}\big[\frac{\epsilon_\nu(\mathbf{P})-\mu_\nu}{k_{\rm B}T}\big]\pm1}.
\label{E:fLNeq}
\end{equation}
In the above, the plus [minus] sign is for fermions [bosons], $H(|\mathbf{P}|-P_\nu^{\rm Mot})$ is the Heaviside step function, and $\mu_\nu$ is the light-cluster chemical potential, which is related to the proton and neutron chemical potentials $\mu_p$ and $\mu_n$ by the chemical equilibrium condition $\mu_\nu$ $=$ $N\mu_n+Z\mu_p$, with $N$ and $Z$ denoting, respectively, the neutron and proton numbers in the light cluster.
The presence of light clusters modifies the occupation of unbound nucleons in nuclear matter $f^{\rm eq}_\tau(\mathbf{p})$ to a form deviating from the Fermi-Dirac distribution. In a nuclear matter of baryon density $\rho_B$, isospin asymmetry $\delta$, and temperature $T$, the quantities $\mu_n$, $\mu_p$, $P_\nu^{\rm Mott}$ in $f^{\rm eq}_\tau(\mathbf{p})$ and $f^{\rm eq}_\nu(\mathbf{P})$ can be determined by solving coupled equations including Eqs.~(\ref{E:fcut}) and (\ref{E:fLNeq}) as described in the supplemental material~\cite{SM} and detailed in Ref.~\cite{WRaXv2506}.

Through Eq.~(\ref{E:fcut}), the Mott effect manifests itself in both nuclear matter and heavy-ion collisions, with its strength characterized by $\mathbf{F}^{\rm cut}$ $\equiv$ $(F^{\rm cut}_2,F^{\rm cut}_3,F^{\rm cut}_4)$.
Calibrating $\mathbf{F}^{\rm cut}$ from the measured light-nuclei yields in heavy-ion collisions based on the kinetic approach allows one to obtain an estimate of the strength of the Mott effects. The preferred $\mathbf{F}^{\rm cut}$ can then be used to deduce the light-cluster fraction in nuclear matter within the phase-space excluded-volume approach. We have checked that the nucleon and light-cluster distributions for nuclear matter obtained from the kinetic approach in a box with periodic boundary conditions are consistent with the analytical $f^{\rm eq}_\tau(\mathbf{p})$ and $f^{\rm eq}_\nu(\mathbf{P})$ (see the Supplemental Material for details), which ensures the viability of the above deduction method.

\emph{Bayesian inference of $\mathbf{F}^{\rm cut}$ from light-nuclei yields.}{\bf ---}
Transport model studies of deuteron and triton production~\cite{KuhPRC63} and $\alpha$-particle production~\cite{WRPRC108} have demonstrated that the Mott effect or the value of $\mathbf{F}^{\rm cut}$ affects appreciably the light-nuclei yields in intermediate-energy heavy-ion collisions. Here, we calibrate the value of $\mathbf{F}^{\rm cut}$ using the light-nuclei yields from these collisions through Bayesian inference based on a Gaussian process emulator of the kinetic approach.
Specifically, using the Surmise package for Bayesian inference~\cite{surmise2021}, we obtain the posterior probability distribution $p(\mathbf{F}^{\rm cut}|\bm{\mathcal{N}}^{\rm exp},\mathcal{K})$ for the $\mathbf{F}^{\rm cut}$ used in the kinetic approach $\mathcal{K}$, given the experimental light-nuclei yields $\bm{\mathcal{N}}^{\rm exp}$ from Au+Au collisions measured by the FOPI Collaboration~\cite{ReiNPA848} (see the Supplemental Material for details). In particular, only collisions at $0.25~\rm{GeV}$, $0.4~\rm{GeV}$, and $0.6A~\rm{GeV}$ are considered in order to minimize the production of heavy fragments at lower collision energies and the contribution of pion-catalyzed reactions for light-nuclei production (e.g., $\pi NN$ $\leftrightarrow$ $\pi d$~\cite{OliPRC99,CocPRC108,SKJNC15}) at higher collision  energies.
This then leads to 12 elements in $\bm{\mathcal{N}}^{\rm exp}$, namely the yields of $d$, $t$, $h$ and $\alpha$ for the above three incident energies.

\begin{figure}[ht]
\centering
\includegraphics[width=\linewidth]{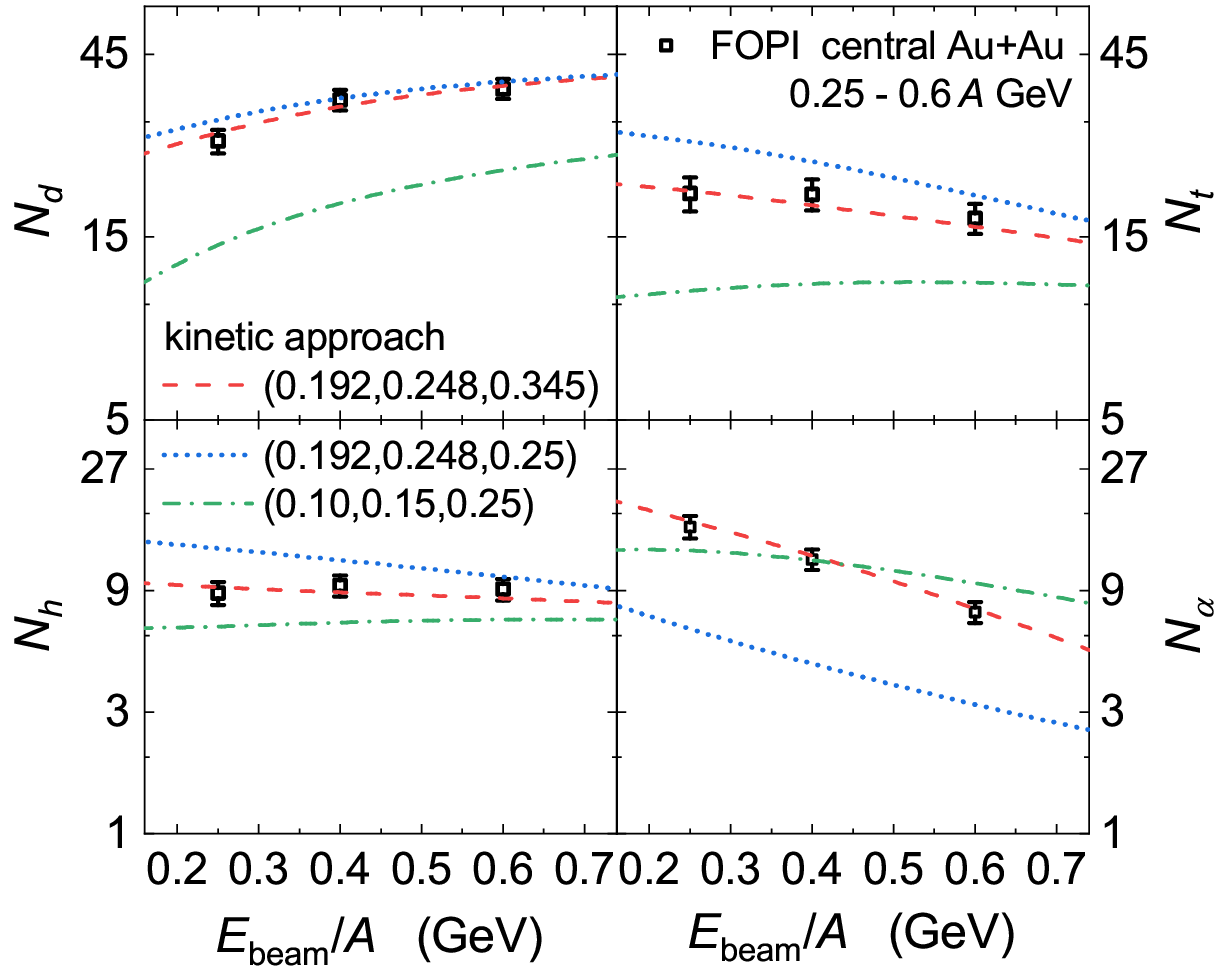}
\caption{Light-nuclei yields in central Au+Au collisions at energies of $0.25$ to $0.6A~\rm GeV$ from the kinetic approach with the most probable value $\mathbf{F}^{\rm cut}=(0.192,0.248,0.345)$, and two other values for $\mathbf{F}^{\rm cut}$.
The experimental data are from the FOPI Collaboration~\cite{ReiNPA848}.}
\label{F:yld}
\end{figure}

At $90\%$ confidence level, the inferred values are $F^{\rm{cut}}_2$ $=$ $0.192_{-0.017}^{+0.037}$ for $d$, $F^{\rm{cut}}_3$ $=$ $0.248_{-0.021}^{+0.075}$ for $t$ and $h$, and $F^{\rm{cut}}_4$ $=$ $0.345_{-0.032}^{+0.102}$ for $\alpha$. Note their differences from those used in Ref.~\cite{WRPRC108}, where only the contribution of unbound nucleons to $f^{\rm tot}(\mathbf{p})$ in Eq.~(\ref{E:fcut}) is considered.
With the preferred value for $\mathbf{F}^{\rm cut}$ $=$ $(0.192,0.248,0.345)$, we show in Fig.~\ref{F:yld} the light-nuclei yields in central Au+Au collisions at the aforementioned three beam energies calculated using the kinetic approach.
Also shown are the results from the other two $\mathbf{F}^{\rm cut}$, namely, $(0.192,0.248,0.25)$ and $(0.10,0.15,0.25)$. Note the interplay between different $F_A^{\rm cut}$, indicating the necessity of employing the Bayesian analysis.


\emph{Alpha-particle fraction in warm and dense nuclear matter.}{\bf ---}
Due to the rapid expansion of the nuclear matter formed in heavy-ion collisions, light clusters can freeze out during the early compression stage of the reaction, and carry information about the properties of the warm and dense (locally equilibrated) nuclear medium where they are emitted.
The preferred values of $\mathbf{F}^{\rm cut}$, determined from the final-state light-nuclei yields, provide an estimate of the strength of the Mott effect and can be used to deduce the light-cluster fractions in such matter.

\begin{figure}[htb]
\centering
\includegraphics[width=5.8cm]{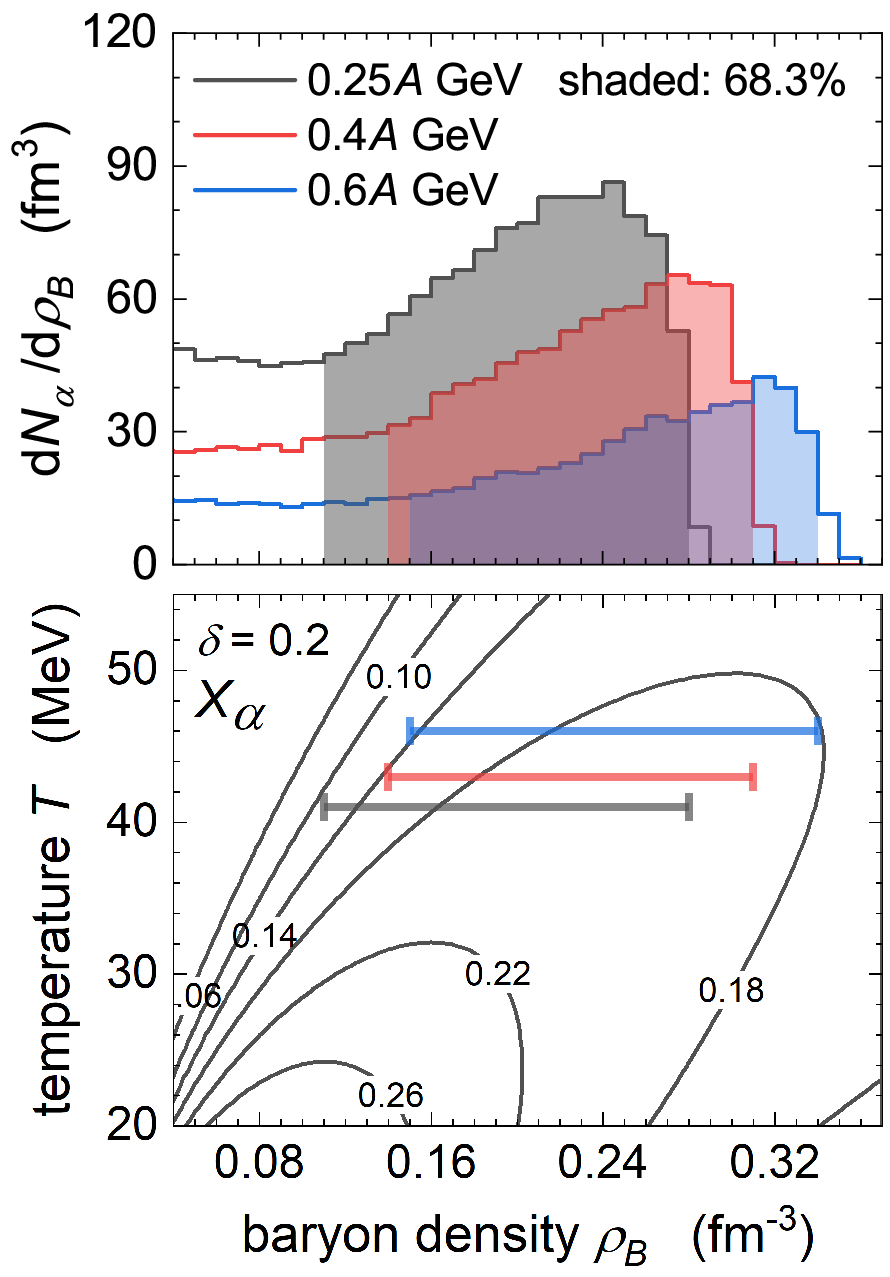}
\caption{Upper: Freeze-out baryon density distribution ${\rm d}N_\alpha/{\rm d}\rho_B$ of $\alpha$-particle number in central Au+Au collisions at $E_{\rm Beam}=0.25A~\rm GeV$, $0.4A~\rm GeV$, and $0.6A~\rm GeV$, obtained from the kinetic approach with the most probable $\mathbf{F}^{\rm cut}$. Shaded regions correspond to the most probable densities within $1\sigma$ confidence level. Bottom: Contour plot of the $\alpha$-particle fraction $X_{\alpha}$ for nuclear matter with isospin asymmetry $\delta$ $=$ $0.2$ obtained using the phase-space excluded-volume approach with the most probable $\mathbf{F}^{\rm cut}$. Error bars indicate the densities of the shaded regions in the upper panel.}
\label{F:LND}
\end{figure}

In the upper panel of Fig.~\ref{F:LND}, we exhibit the distribution of the local baryon density $\rho_B$ associated with the number of chemical freeze-out $\alpha$ particles in heavy-ion collisions, as extracted using the kinetic approach. $\alpha$ particles are seen to mainly freeze out chemically from high-density regions during the most compressed stage of the collisions. This is in contrast to deuterons, tritons and helium-$3$ that mainly freeze out from lower densities because of their less enhanced formation rate in dense nuclear matter due to the $\rho_N^{A+1}$ dependence of many-body scatterings employed in the kinetic approach for their production. Therefore, we focus on the $\alpha$ particle and deduce its fraction in warm and dense nuclear matter. Specifically, the chemically freeze-out $\alpha$-particle number distribution ${\rm d}N_\alpha/{\rm d}\rho_B$ is found to peak around $\rho_B$ $=$ $0.25_{-0.14}^{+0.03}~\rm fm^{-3}$, $0.28_{-0.14}^{+0.03}~\rm fm^{-3}$, and $0.32_{-0.17}^{+0.02}~\rm fm^{-3}$ for $E_{\rm beam}$ $=$ $0.25A~\rm GeV$, $0.4A~\rm GeV$, and $0.6A~\rm GeV$, respectively, where the lower and upper bounds correspond to the $1\sigma$ confidence interval of the distribution.

To avoid ambiguities in determining local temperatures of nuclear matter within the kinetic approach, we extrapolate the chemical freeze-out temperatures determined from the statistical hadronization model used for describing hadron yields in high-energy nuclear collisions~\cite{AndNt561} down to intermediate energies. The estimated chemical freeze-out temperatures around which nucleons freeze-out in Au+Au collisions at $E_{\rm beam}$ $=$ $0.25A~\rm GeV$, $0.4A~\rm GeV$, and $0.6A~\rm GeV$, are found to be $41~\rm MeV$, $43~\rm MeV$, and $46~\rm MeV$, respectively.

With the density $\rho_\nu$ of light clusters of certain species obtained by integrating its phase-space distribution in Eq.~(\ref{E:fLNeq}), its fraction in nuclear matter $X_\nu=\frac{A\rho_\nu}{\rho_B}$ can be determined.
We show in the bottom panel of Fig.~\ref{F:LND} the contour plot of the $\alpha$-particle fraction $X_\alpha$ in nuclear matter with isospin asymmetry $\delta$ $=$ $0.2$, which roughly corresponds to the isospin asymmetry of the Au+Au collision system, around the aforementioned chemical freeze-out densities and temperatures in the $\rho_B$--$T$ phase diagram, obtained using the phase-space excluded-volume approach with the most probable $\mathbf{F}^{\rm cut}$ $=$ $(0.192,0.248,0.345)$.
The error bars correspond to the freeze-out density intervals shown in the upper panel.
A surprisingly high $\alpha$-particle fraction of $X_{\alpha}$ $\sim$ $0.2$ at the chemical freeze-out densities of $0.25$--$0.32~\rm fm^{-3}$ and temperatures of $41$--$46$ MeV is observed.
It indicates that the picture of warm and dense nuclear matter as a uniform nucleon liquid is inconsistent with the abundant light-nuclei yields observed in intermediate-energy heavy-ion collisions.
This result will have crucial implications on the nuclear equation of state, and the physics of supernovae and neuron star mergers. 

\begin{figure}[!h]
\centering
\includegraphics[width=5.8cm]{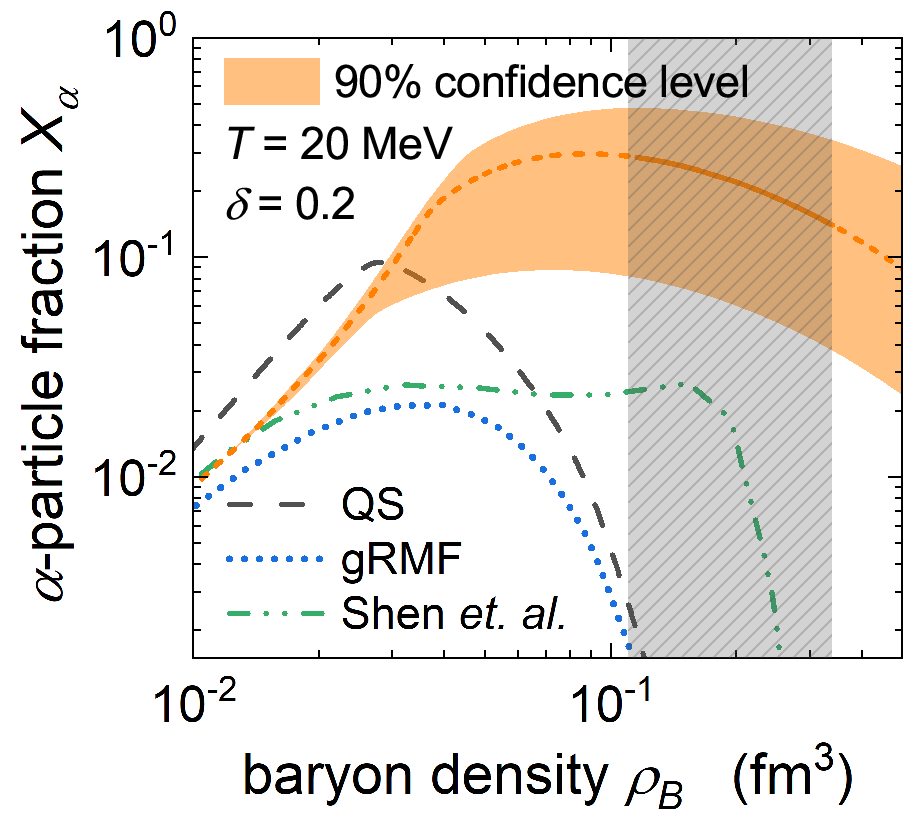}
\caption{$\alpha$-particle fraction $X_{\alpha}$ in nuclear matter at temperature $T$ $=$ $20~\rm MeV$ and isospin asymmetry $\delta$ $=$ $0.2$ obtained from the phase-space excluded-volume approach with the most probable $\mathbf{F}^{\rm cut}$ $=$ $(0.192,0.248,0.345)$. Its uncertainty are obtained by varying $F_2^{\rm cut}$,  $F_3^{\rm cut}$ and $F_4^{\rm cut}$ within their $90\%$ confidence intervals. The results from QS, gRMF, and Shen {\it et~al.} (for $\delta$ $=$ $0$)~\cite{TypPRC81} are included for comparison. The highlighted area corresponds to the dominant density region where $\alpha$ particles undergo the chemical freeze out shown in Fig.~\ref{F:LND}.}
\label{F:Xa}
\end{figure}

We compare in Fig.~\ref{F:Xa} the above results on the $\alpha$-particle fraction $X_{\alpha}$ in nuclear matter of $\delta=0.2$ and at $T=20$ MeV with those from other approaches.
These approaches (Their results are available only for $\delta$ $=$ $0$. Note that $X_\alpha$ is generally suppressed in asymmetric matter.) include the generalized relativistic mean-field (gRMF) model~\cite{TypPRC81}, where the Mott effect is implemented effectively by introducing certain density-dependent (momentum-independent) binding energies of light clusters, and the Thomas-Fermi approximation with the dissolution of $\alpha$ particles realized by an excluded-volume prescription in coordinate space (Shen {\it et~al.})~\cite{SHNPA637}.
The $X_\alpha$ obtained from the quantum statistical (QS) approach, where the Mott effect is implemented directly by parameterizing the in-medium binding energies of light clusters from the solution of the in-medium Schrodinger equation, is also included as a reference. Note that in this approach, light-cluster fractions contain contributions from both bound-state correlations (light clusters discussed in this work) and scattering-state correlations.
The orange line represents $X_{\alpha}$ from the phase-space excluded-volume approach [Eq.~(\ref{E:fcut})] using the most probable $\mathbf{F}^{\rm cut}$, with its uncertainty obtained by varying $F_2^{\rm cut}$, $F_3^{\rm cut}$ and $F_4^{\rm cut}$ within their $90\%$ confidence intervals.
The highlighted area corresponds to the dominant density region (combining the regions for the three $E_{\rm beam}$ together) where $\alpha$ particles undergo the chemical freeze-out shown in Fig.~\ref{F:LND}. It is seen that the $X_{\alpha}$ at these densities preferred by the light-nuclei yields in intermediate-energy heavy-ion collisions is much larger than that predicted by previous approaches.
A much weaker Mott effect on light clusters in warm and dense nuclear matter needs to be introduced in those approaches to make their results agree with the light-nuclei yields measured by the FOPI collaboration. We notice that the in-medium effects on deuterons have already been modified within the gRMF approach~\cite{BurEPJA58} to explain the short-range correlations observed in finite nuclei~\cite{HenRMP89}.

\emph{Summary.}{\bf ---}
Based on a kinetic approach that dynamically includes light clusters with mass number $A$ $\leqslant$ $4$ and the Mott effect, we are able to study the in-medium effects on light clusters in warm and dense nuclear matter--—previously possible only for dilute matter.
Our study demonstrates that the conventional view of dense nuclear matter as a uniform nucleon liquid is inconsistent with the light-nuclei yields measured in central Au+Au collisions at intermediate energies by the FOPI collaboration.
According to the measured yields, an unexpectedly abundant $\alpha$ clustering with a fraction of $\sim$ $0.2$ should exist in warm nuclear matter at densities around $1$-$2$ times of nuclear saturation density ($\rho_0$ $\approx$ $0.16~\rm fm^{-3}$). 
Such a high $\alpha$-particle fraction has a profound impact on our understanding of the equation of state of warm and dense nuclear matter.
It underscores the need to extend nuclear mean-field models to explicitly include light-cluster degrees of freedom and the Mott effect, as has been done in the quantum statistical approach and the generalized relativistic mean-field model~\cite{TypPRC81,ZZWPRC95}.
Our results also have important implications for various astrophysical phenomena, such as supernovae and binary neutron star mergers, where similar conditions can arise.

\begin{acknowledgments}
~~

We thank Xin Li and Chen Zhong for the assistance when using the GPU server. Rui Wang thanks Stefano Burrello, Maria Colonna, and Stefan Typel for useful discussion. This work was supported in part by the National Natural Science Foundation of China under contract Nos. $12147101$, $12235010$, $12422509$, and $12375121$, the National Key Research and Development Program of China under Grant Nos. $2024$YFA$1612500$, and $2018$YFE$0104600$, the National SKA Program of China No. $2020$SKA$0120300$, the Guangdong Major Project of Basic and Applied Basic Research No. $2020$B$0301030008$, the Science and Technology
Commission of Shanghai Municipality under Grant No. $23$JC$1402700$, $23$JC$1400200$, and $23590780100$, and the U.S. Department of Energy under Award No. DE-SC$0015266$.
\end{acknowledgments}

\bibliography{LN_Mott}

\end{document}


\title{Supplemental Material for ``Alpha clustering in warm and dense nuclear matter from heavy-ion collisions"}

\date{\today}

\maketitle

\section{Nuclear matter in the phase-space excluded-volume approach}
\label{S:NM}

In order to compare the phase-space occupation of nucleons (including neutrons $n$ and protons $p$) and light clusters in nuclear matter obtained with the kinetic approach in a box with the corresponding analytical results, we first outline how the properties of nuclear matter are obtained within the phase-space excluded-volume approach.

In nuclear matter with light clusters, from the collision integral of the BUU equation, the momentum occupation of unbound nucleons can be expressed as~\cite{WRaXv2506}, 
\begin{equation}
    f^{\rm eq}_{\tau}(\vec{p}) = \frac{1-f_{\tau}^{\rm clu}(\vec{p})}{{\rm exp}\big\{\beta[\epsilon_{\tau}(\vec{p})-\mu_{\tau}]\big\} + 1},
\label{E:feq}
\end{equation}
with $\tau$ $=$ $n$ or $p$ and $\beta$ being the inverse of temperature $T$. $\epsilon_{\tau}(\vec{p})$ and $\mu_\tau$ denote, respectively, the single particle energy and the nucleon chemical potential. The $f_{\tau}^{\rm clu}(\vec{p})$ represents the nucleon occupation originating solely from  nucleons bound in light clusters. It depends on the light-cluster phase-space occupation $f_\nu(\vec{P})$ and the internal wave function of the light clusters, i.e. (see also Ref.~\cite{RopPRC79}), 
\begin{eqnarray}
    f^{\rm clu}_n(\vec{p}) = \sum_{\nu}^{\rm clu}N\int f_{\nu}(\vec{P})|\phi_{\nu,\vec{P}}(\vec{p})|^2{\rm d}\vec{P}\label{E:fnclu},\\
    f^{\rm clu}_p(\vec{p}) = \sum_{\nu}^{\rm clu}Z\int f_{\nu}(\vec{P})|\phi_{\nu,\vec{P}}(\vec{p})|^2{\rm d}\vec{P}\label{E:fpclu}.
\end{eqnarray}
In the above, the summation runs over $\nu$ $=$ $d$ (deuterons), $t$ (tritons), $h$ (helium-$3$) and $\alpha$ ($\alpha$ particles), while $N$ and $Z$ denote the neutron and proton numbers of the light-cluster species $\nu$, respectively. The $|\phi_{\nu,\vec{P}}(\vec{p})|^2$ represents the normalized one-body probability distribution of nucleons bound in light clusters of species $\nu$ and center-of-mass momentum $\vec{P}$. For simplicity, we do not distinguish between neutrons and protons and approximate its value in free space using a Gaussian function of the form, 
\begin{equation}
    |\phi_{\nu,\vec{P}}(\vec{p})|^2 \propto {\rm exp}\Big[-\frac{2\big(\vec{p}-\frac{\vec{P}}{A}\big)^2}{\hbar^2}\sigma^2_{\nu}\Big].
\label{E:Gau}
\end{equation}
The $\sigma_\nu$ is determined by the r.m.s. radius of light nuclei in free space, namely, $\sigma_d$, $\sigma_t$, $\sigma_h$ and $\sigma_\alpha$ $=$ $2.26~\rm fm$, $1,59~\rm fm$, $1.76~\rm fm$, and $1.54~\rm fm$, respectively.  The same parameters are also adopted when solving the kinetic equations numerically, although they could depend on the phase-space density of the nuclear medium.
However, the difference between the in-medium $|\phi_{\nu,\vec{P}}(\vec{p})|^2$ and its free-space counterpart is found to be small in a study based on the variational calculation~\cite{YSPRC108}.

The light-cluster phase-space occupation $f_\nu(\vec{P})$ in nuclear matter follows the Fermi-Dirac or Bose-Einstein distribution, as it can still be obtained from the collision integral of the BUU equation, which includes many-body scatterings of light clusters, such as $nnp$ $\leftrightarrow$ $nd$. However, the $f_\nu(\vec{P})$ is modified by the Mott effect, i.e., a light cluster can exist only if the magnitude of its momentum is greater than its Mott momentum $P^{\rm Mott}_\nu$, to the following form, 
\begin{equation}
\begin{split}
    f^{\rm eq}_{\nu}(\vec{P}) = \frac{ H\big(|\vec{P}|-P^{\rm Mott}_{\nu}\big)}{{\rm exp}\big\{\beta[\epsilon_{\nu}(\vec{P})-\mu_{\nu}]\big\}\pm1},
\end{split}
\label{E:fLNeq}
\end{equation}
where the plus [minus] sign is for fermions [bosons] and $H$ is the Heaviside step function. The Mott momentum $P^{\rm Mott}_{\nu}$ can be obtained from the phase-space excluded-volume approach, i.e., Eq.~(3) in the main article, and it is given by~\footnote{This can be realized by noting that in nuclear matter, $f_{\tau}^{\rm tot}(\vec{p})$ decreases monotonously as $|\vec{p}|$ increases, and consequently, $\langle f_{\tau} \rangle_{\nu}(\vec{P})$ also decreases with increasing $|\vec{P}|$. However, if $P^{\rm Mott}_\nu$ is sufficiently large, $f^{\rm clu}_\tau(\vec{p})$ exhibits a peak around $P^{\rm Mott}_\nu/A$, and it is possible that $f^{\rm tot}_\tau(\vec{p})$ becomes non-monotonic.
This scenario is omitted as it arises only for unreasonably large $F^{\rm cut}_A$.}
\begin{equation}
    \begin{cases}
    P^{\rm Mott}_{\nu} = 0,~~~~~~{\rm if}~\langle f_\tau \rangle_\nu(\vec{P}=0)<F^{\rm cut}_A,~\tau = n,p,\\
    \max\limits_{\tau=n,p}\big[\langle f_\tau\rangle_\nu(P^{\rm Mott}_{\nu})\big] = F^{\rm cut}_A,~~~~~~~~~~{\rm if}~P^{\rm Mott}_{\nu} \ne 0. 
    \end{cases}
\label{E:PMt}
\end{equation}
The chemical potential of light clusters in nuclear matter is determined by the condition of chemical equilibrium, i.e., $\mu_\nu$ $=$ $N\mu_n$ $+$ $Z\mu_p$.

Adding $f_{\tau}(\vec{p})$ with $f^{\rm clu}_{\tau}(\vec{p})$ leads to the total nucleon occupation,
\begin{equation}
\begin{split}
    f^{\rm tot,eq}_{\tau}(\vec{p}) \equiv f^{\rm eq}_{\tau}(\vec{p}) + f^{\rm clu}_{\tau}(\vec{p}) = \frac{1+f^{\rm clu}_{\tau}(\vec{p}){\rm exp}\big\{\beta[\epsilon_{\tau}(\vec{p})-\mu_{\tau}]\big\}}{{\rm exp}\big\{\beta[\epsilon_{\tau}(\vec{p})-\mu_{\tau}]\big\} + 1}.
\label{E:ftotNM}
\end{split}
\end{equation}
The $f^{\rm clu}_{\tau}(\vec{p})$ in the above equation is obtained by substituting the $f_\nu(\vec{P})$ in Eqs.~(\ref{E:fnclu}) and (\ref{E:fpclu}) into the light-cluster occupations at equilibrium $f^{\rm eq}_\nu(\vec{P})$ given in Eq.~(\ref{E:fLNeq}).  For a given temperature $T$ and total nucleon density $\rho^{\rm tot}_\tau$, the thermal properties of a nuclear matter can then be determined by adjusting the chemical potential $\mu_\tau$ in $f^{\rm tot}_{\tau}(\vec{p})$ to fulfill the condition,
\begin{equation}
g \int \frac{{\rm d}\vec{p}}{(2\pi\hbar)^3} f^{\rm tot}_{\tau}(\vec{p}) = \rho_{\tau}^{\rm tot},
\label{E:NM}
\end{equation}
with $g$ being the nucleon spin degeneracy. This involves solving the coupled Eqs.~(\ref{E:fnclu}), (\ref{E:fpclu}), and (\ref{E:fLNeq})--(\ref{E:NM})  simultaneously, as $f^{\rm tot}_{\tau}(\vec{p})$ depends on itself through the dependence of $f^{\rm clu}_{\tau}(\vec{p})$ on the Mott momentum of light clusters. Details of the in-medium properties of light clusters obtained from the phase-space excluded-volume approach can be found in Ref.~\cite{WRaXv2506}.


\section{Box calculation using the kinetic approach}

In addition to the above analytical expressions, the properties of nuclear matter, such as particle fractions and occupations in momentum space [$f^{\rm eq}_\tau(\vec{p})$ in Eq.~(\ref{E:feq}) and $f^{\rm eq}_\nu(\vec{P})$ in Eq.~(\ref{E:fLNeq})], can also be obtained using the kinetic approach. We perform simulations of the time evolution of nucleons and light clusters confined in a box with periodic boundary conditions  using the kinetic approach. We initialize a system confined in a $10\times10\times10~\rm fm^3$ box that consists of an equal number of $80$ neutrons and protons, corresponding to $\rho_B$ $=$ $0.16~\rm fm^3$ $\sim$ $\rho_0$.  Due to the inclusion of many-body scatterings, nucleons and light clusters convert into each other as the system evolves, eventually reaching thermal equilibrium. The initial momenta of neutrons and protons follow the Fermi-Dirac distribution with a temperature of $T$ $=$ $17.55~\rm MeV$. After equilibrium, the system results in clustered nuclear matter at $T$ $=$ $20~\rm MeV$, driven by energy conservation.

\begin{figure}[htp]
\centering
\includegraphics[width=0.5\linewidth]{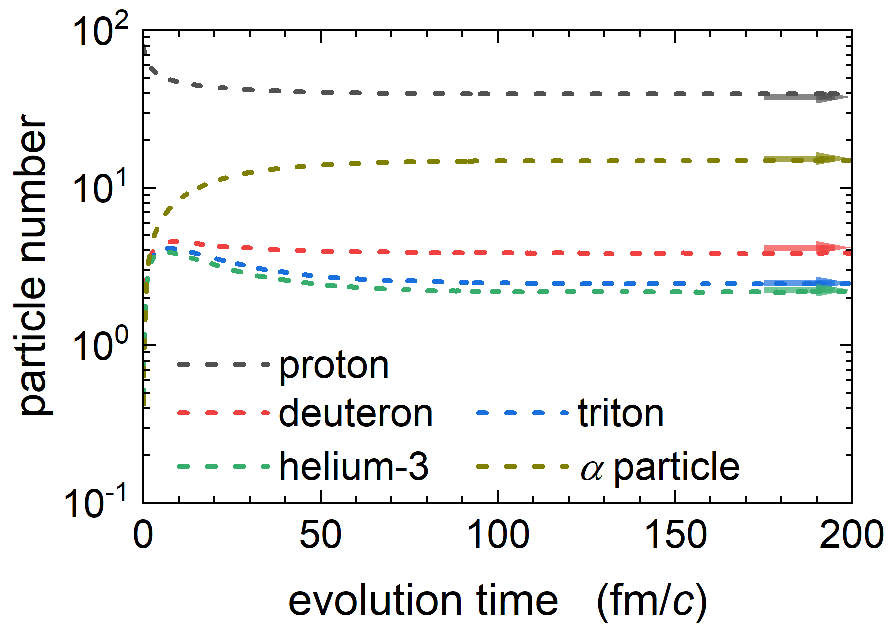}
\caption{Time evolution of proton, deuteron, triton, helium-3 and $\alpha$-particle numbers in the box calculation using the kinetic approach. The number of ensemble $N_{\rm E}$ (test-particle number) used in the calculation is $2^{14}$. Arrows indicate the equilibrium particle numbers obtained with the analytical expressions given in Sec.~\ref{S:NM} for for total neutron and proton densities $\rho^{\rm tot}_n$ $=$ $\rho^{\rm tot}_p$ $=$ $0.08~\rm fm^{-3}$ and temperature $T$ $=$ $20~\rm MeV$.}
\label{F:yldbox}
\end{figure}

In Fig.~\ref{F:yldbox}, we present the time evolution of particle numbers obtained using the kinetic approach in the above box calculations.  The arrows indicate their thermal equilibrium values for symmetric nuclear matter at baryon density $\rho_B$ $=$ $0.16~\rm fm^{-3}$ and temperature $T$ $=$ $20~\rm MeV$ calculated employing the analytical formulae in Sec.~\ref{S:NM}. In both calculations, we adopt the cutoff parameters ${\bf F}^{\rm cut}$ $=$ $(0.2,0.25,0.35)$. As seen from the figure, the equilibrium particle numbers obtained using the kinetic approach are consistent with corresponding analytical values.

\begin{figure}[htp]
\centering
\includegraphics[width=0.55\linewidth]{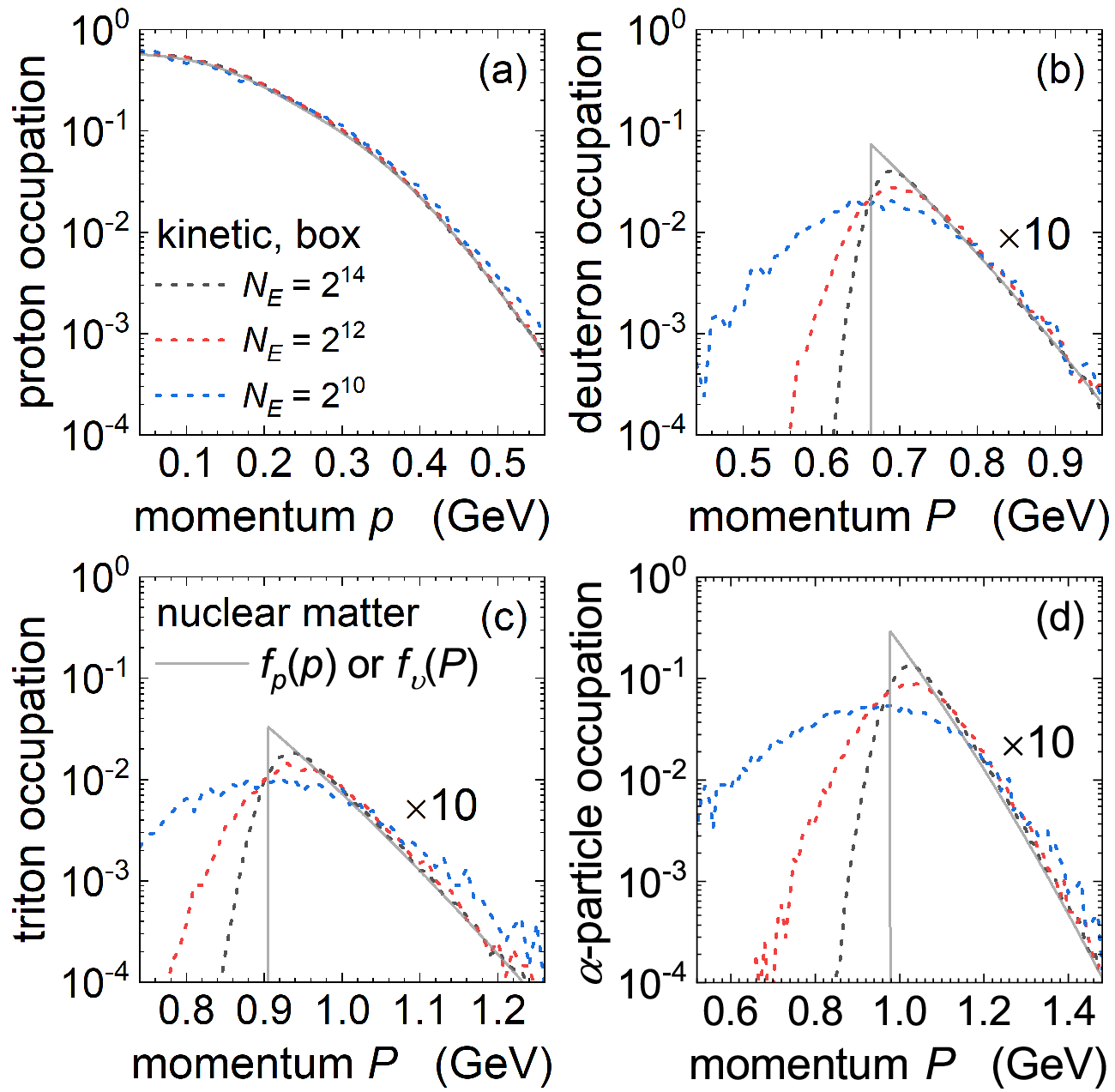}
\caption{Occupations in momentum space for protons (a), deuterons (b), tritons (c), and $\alpha$-particles (d) after the box reaches equilibrium, obtained within the kinetic approach using different numbers of ensembles $N_{\rm E}$. The corresponding analytical occupations in symmetric nuclear matter at baryon density $\rho_B$ $=$ $0.16~\rm fm^{-3}$ and temperature $T$ $=$ $20~\rm MeV$, calculated with the phase-space excluded-volume approach [Eqs.~(\ref{E:feq}) and (\ref{E:fLNeq})] are also shown for comparison. In both calculations, the cutoff parameter set ${\bf F}^{\rm cut}$ is chosen as $(0.2,0.25,0.35)$.}
\label{F:f}
\end{figure}

To further validate the capability of the kinetic approach in describing the properties at thermal equilibrium, we present in Fig.~\ref{F:f} the phase-space occupations of protons, deuterons, tritons, and $\alpha$-particles obtained from the kinetic approach after the box reaches equilibrium. For reference, the analytical occupations of these particle species, calculated using the phase-space excluded-volume approach, are shown as gray solid lines. Note the sharp edges in the deuteron, triton, and $\alpha$-particle occupations shown in window (b), (c), and (d), respectively. These arise from the Heaviside function in Eq.~(\ref{E:fLNeq}), which reflects their corresponding Mott momenta. As the number of ensembles $N_{\rm E}$ (or number of test particles) used in the kinetic approach increases, the resulting light-cluster occupations approach their analytical counterparts, with low-momentum edges becoming increasingly sharper. However, due to fluctuations in the numerical calculations of $\langle f_\tau\rangle_\nu(\vec{P})$  within the kinetic approach, a perfectly sharp edge at the Mott momentum can only be achieved in the limit of infinite $N_{\rm E}$. When evaluating the light-nuclei yields in heavy-ion collisions using the kinetic approach, we adopt a sufficiently large $N_{\rm E}$ so that further increases produce negligible changes in the obtained particle yields.

\section{Bayesian inference of $\mathbf{F}^{\rm cut}$ from light-nuclei yields}

\begin{figure}[tp]
\centering
\includegraphics[width=0.55\linewidth]{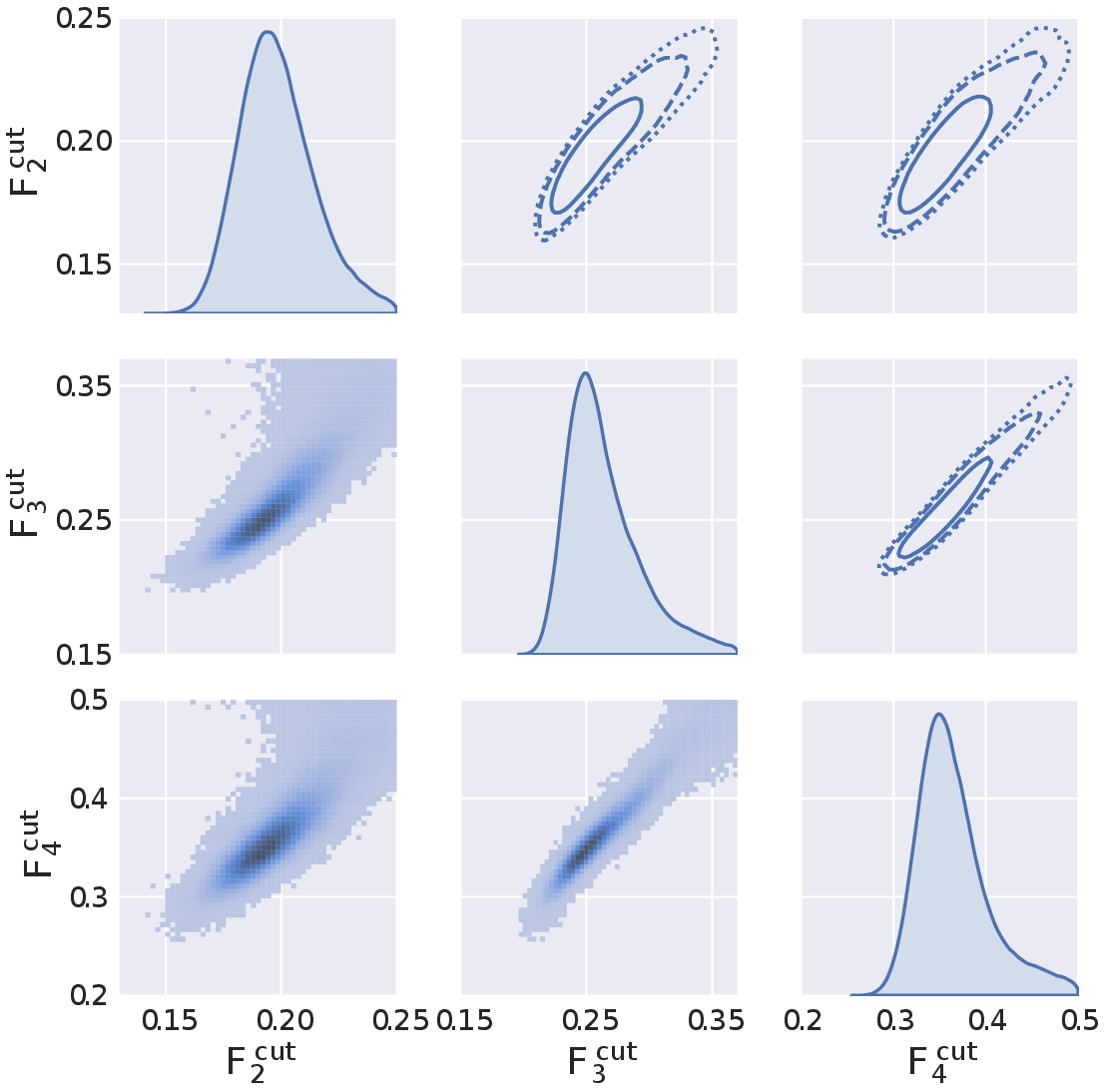}
\caption{Posterior univariate (diagonal panels) and bivariate (off-diagonal panels) distributions of $F^{\rm cut}_A$. The solid, dashed, and dotted contours in the upper-right off-diagonal panels correspond to the $68.3\%$, $90.0\%$, and $95.5\%$ confidence intervals, respectively.}
\label{F:post}
\end{figure}

\begin{figure}[!b]
\centering
\includegraphics[width=0.55\linewidth]{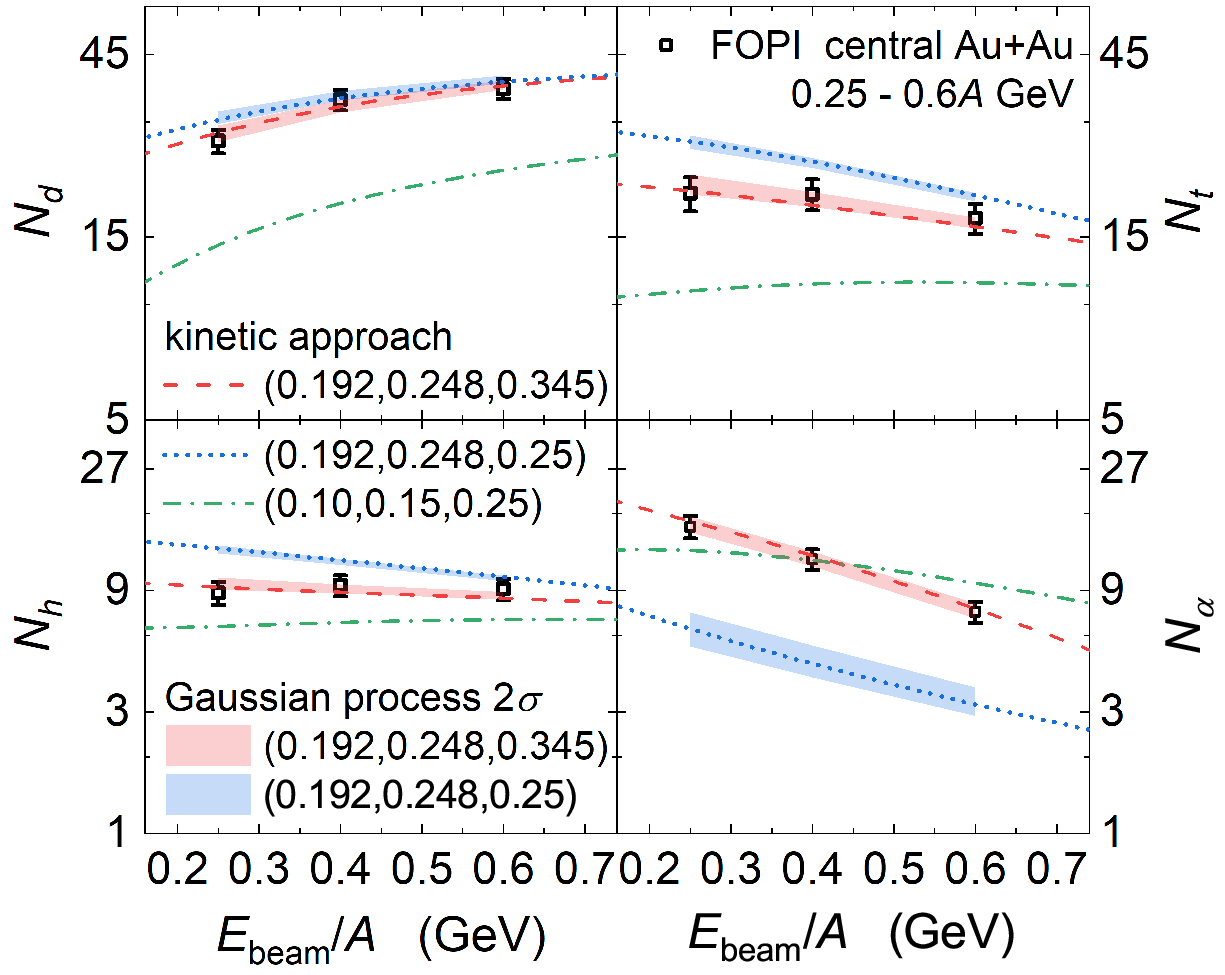}
\caption{Same as the Fig.~$1$ in the main article, but with the inclusion of Gaussian process predictions for ${\bf F}^{\rm cut}$ $=$ $(0.192,0.248,0.345)$ and $(0.192,0.248,0.25)$. The bands represents the associated uncertainties at the  $2\sigma$ confidence level.}
\label{F:yld}
\end{figure}

This section provides additional details of and discussion on the Bayesian inference of ${\bf F}^{\rm cut}$. As mentioned in the main article, using the Surmise package for Bayesian inference~\cite{surmise2021}, we obtain the posterior probability distribution $p(\mathbf{F}^{\rm cut}|\bm{\mathcal{N}}^{\rm exp},\mathcal{K})$ for the cutoff parameter set $\mathbf{F}^{\rm cut}$ used in the kinetic approach $\mathcal{K}$, given the experimental light-nuclei yields $\bm{\mathcal{N}}^{\rm exp}$ from Au+Au collisions at beam energies $E_{\rm beam}$ $=$ $0.25A~\rm GeV$, $0.4A~\rm GeV$, and $0.6A~\rm GeV$ measured by the FOPI Collaboration~\cite{ReiNPA848}. The vector $\bm{\mathcal{N}}^{\rm exp}$ contains 12 elements, namely the yields of deuterons ($d$), tritons ($t$), helium-3 ($h$), and $\alpha$ particles at the above three incident energies. The priors $p({\bf F}^{\rm cut})$ for the cutoff parameters $F^{\rm cut}_2$, $F^{\rm cut}_3$, and $F^{\rm cut}_4$ are chosen as uniform distributions over the ranges $0.13$--$0.25$, $0.15$--$0.37$, and $0.20$--$0.50$, respectively. Since a sufficiently large number of ensembles $N_{\rm E}$ is required to ensure convergence of the light-nuclei yields (as discussed in the previous section), solving the kinetic equations for a given ${\bf F}^{\rm cut}$ typically takes a few days. This makes direct application of the kinetic approach in the Bayesian analysis impractical and inefficient.  To overcome this obstacle, a Gaussian process emulator of the kinetic approach is employed. Thirty sets of ${\bf F}^{\rm{cut}}$ are sampled within their prior ranges using Latin hypercube sampling. The kinetic approach results for these parameter sets serve as training points for the Gaussian process emulator. Assuming a Gaussian likelihood, the posterior distribution is then estimated via Metropolis-Hastings sampling based on the constructed emulator.
The obtained univariate (diagonal panels) and bivariate (off-diagonal panels) distributions of the posterior $p(\mathbf{F}^{\rm cut}|\bm{\mathcal{N}}^{\rm exp},\mathcal{K})$ are shown in Fig.~\ref{F:post}. The solid, dashed, and dotted contours in the upper-right off-diagonal panels of the figure correspond to the $68.3\%$, $90.0\%$, and $95.5\%$ confidence regions, respectively.

To validate the Gaussian process emulator, we choose two cutoff parameter sets ${\bf F}^{\rm cut}$, namely, $(0.192,0.248,0.345)$ and $(0.192,0.248,0.25)$.  Figure~\ref{F:yld} reproduces Fig.~$1$ in the main article obtained using the kinetic approach.
In Fig.~\ref{F:yld}, in addition to the results obtained from the kinetic approach using the two ${\bf F}^{\rm{cut}}$ sets, the $2\sigma$ confidence intervals predicted by the Gaussian process emulator are shown as shaded bands. As seen in the figure, both the deviations of the emulator predictions from the actual model calculations and their associated uncertainties are significantly smaller than the 
experimental uncertainties.
This confirms the reliability of the Gaussian process emulator as a fast and accurate surrogate for the full model calculations. 
As for the other ${\bf F}^{\rm cut}$ $=$ $(0.1,0.15,0.25)$, since it lies outside the prior distribution range and is far from the training points, a comparison between the light-nuclei yields predicted by the Gaussian process and those directly from the kinetic approach is not particularly meaningful.

\bibliography{LN_Mott}